\documentclass[aps,prb,reprint,superscriptaddress,longbibliography]{revtex4-1}

\usepackage{graphicx}
\usepackage{latexsym}
\usepackage{amsmath}
\usepackage{amssymb}
\usepackage{amsfonts}
\usepackage{color}
\usepackage{bm}% bold math
\usepackage{verbatim}
\usepackage{hyperref}

\definecolor{MS-color}{RGB}{128,0,128}

\usepackage{framed}
\definecolor{shadecolor}{RGB}{222,222,221}
\begin{document}

\title{Renormalization of antiferromagnetic magnons by superconducting condensate and quasiparticles}

\author{A. M. Bobkov}
\affiliation{Moscow Institute of Physics and Technology, Dolgoprudny, 141700 Moscow region, Russia}

\author{S. A. Sorokin}
\affiliation{National Research University Higher School of Economics, 101000 Moscow, Russia}
\affiliation{Moscow Institute of Physics and Technology, Dolgoprudny, 141700 Moscow region, Russia}

\author{I. V. Bobkova}
\affiliation{Moscow Institute of Physics and Technology, Dolgoprudny, 141700 Moscow region, Russia}
\affiliation{National Research University Higher School of Economics, 101000 Moscow, Russia}

\date{\today}

 %%% abstract

\begin{abstract}
The ability to modify and tune
the spin-wave dispersion is one of the most important
requirements for engineering of magnonic networks. In this study we demonstrate the promise of synthetic thin-film hybrids composed of an antiferromagnetic insulator (AF) and a normal (N) or superconducting (S) metal for tuning and modifying the spin-wave dispersion in antiferromagnetic insulators. The key ingredient is the uncompensated magnetic moment at the AF/S(N) interface, which induces an effective exchange field in the adjacent metal via the interface exchange interaction. The exchange field spin polarizes quasiparticles in the metal and induces spinful triplet Cooper pairs screening the magnon. The quasiparticle and Cooper pair polarization renormalizes the magnon dispersion. The renormalization results in the splitting of the otherwise degenerate AF magnon modes with no need to apply a magnetic field. It is also proposed that measurements of the
renormalized dispersion relations can provide the amplitude of the effective exchange
field induced by the AF in the adjacent metal.
\end{abstract}

 %%% PACS numbers
\maketitle
 
\section{Introduction}

Magnonics is devoted to the exploration of spin waves in magnetic structures. Now it is very rapidly developing mainly due to the impressive advance of nanotechnology, the development of new experimental techniques and the promise of a new generation of devices in which magnons would be employed \cite{Democritov2013,Rezende2020,Barman2021}. In particular, the interconversion between magnonic spin signals and electron spin signals has been investigated \cite{Chumak2015,Kajiwara2010,Kamra2016,Weiler2013,KAto2019,Amudsen2022}. 

The other important question is a renormalization of magnon characteristics in hybrid structures, and, in particular, the influence of the adjacent metal on the magnon characteristics in thin film ferromagnet/normal metal (F/N) or ferromagnet superconductor (F/S) heterostructures. The ability to modify and
tune the spin-wave dispersion is one of the most important requirements for engineering of magnonic networks. In particular, it was found that the adjacent metal works as a spin sink strongly influencing Gilbert damping of the magnon modes \cite{Tserkovnyak2005,Ohnuma2014,Bell2008,Jeon2019,Jeon2019_2,Jeon2019_3,Jeon2018,Yao2018,Li2018,Golovchanskiy2020,Silaev2020}. Also, the adjacent superconducting layer can result in shifting of $k=0$ magnon frequencies (Kittel mode) \cite{Li2018,Golovchanskiy2020,Jeon2019_3,Silaev2020}. The exact mechanism is still under debate now \cite{Silaev2022,Mironov2021}, but it is probably of an electromagnetic nature and is not related to the exchange-based proximity effects between the ferromagnet and the superconductor. The electromagnetic interaction between the ferromagnet and superconductor also results in the appearance of skyrmion-fluxon excitations\cite{Hals2016,Baumard2019,Dahir2019,Menezes2019,Andriyakhina2021,Petrovic2021,Bihlmayer2021}, magnon-fluxon excitations \cite{Dobrovolskiy2019} and efficient gating of magnons\cite{Yu2022}. Further it was reported that the proximity effect in thin film F/S hybrids, which is the generation of an effective exchange field in S by the proximity to F, results in the appearance of composite particles of a different physical nature. They are composed of a magnon in F and an accompanying cloud of spinful triplet pairs in S, which was termed magnon-cooparon \cite{Bobkova2022}. The cloud of triplet pairs results in the essential renormalization of the magnon stiffness and its spin. Furthermore, for thin film S/F bilayers with an unconventional spinful triplet superconductor, a conversion of Cooper pair supercurrents to magnon spin currents has been suggested\cite{Johnsen2021}.

On the other hand, antiferromagnets (AFs) have recently
gathered interest as alternatives to ferromagnets 
as active components in spintronics applications due to  their robustness against perturbation by magnetic fields, the absence
of parasitic stray fields and ultrafast dynamics\cite{Baltz2018,Jungwirth2016,Brataas2020}. In particular, antiferromagnet/normal metal hybrids have been actively studied in the context of spin transport experiments\cite{Wimmer2020,Kamra2017,Cheng2014,Erlandsen2022}. Antiferromagnetic magnons have been predicted to be effective mediators of the superconducting pairing in antiferromagnet/metal hybrid structures \cite{Erlandsen2019,Erlandsen2020,Erlandsen2020_2,Thingstad2021}. 

The influence of superconductivity on the magnon spectra in antiferromagnetic superconductors has been studied in Ref.~\onlinecite{Buzdin1986}. It was reported that the superconductivity enhances $k=0$ magnon frequency thus making the magnon spectrum a nonmonotonic function of the magnon wave number.  However, much less is known about the back action of the adjacent metal on the magnon characteristics in thin film AF/N or AF/S heterostructures. Here we address this question for an antiferromagnetic insulator/normal (superconducting) metal interface if the corresponding interface is uncompensated, that is, there is a nonzero average magnetization at the interface. Due to the interface exchange coupling an effective exchange field is induced in the adjacent metal producing an electron spin polarization there \cite{Kamra2018}. This makes the thin conducting film proximitized by the uncompensated antiferromagnet very much like a conductor in a contact with a ferromagnetic insulator. The effective exchange field  generates singlet-triplet conversion in the superconducting film\cite{Buzdin2005}, superconductivity suppression, spin-splitting of the superconducting DOS\cite{Bergeret2018} and giant spin-dependent Seebeck effect \cite{Bobkov2021}. 

We study the back action of the spin polarization induced in the superconductor or normal metal on the magnon spectrum of the antiferromagnetic insulator. We consider an easy-axis antiferromagnet. It is known that in this case there are two oppositely polarized magnon modes, which are degenerate unless an external magnetic field is applied. We demonstrate that the back action of the metal removes the degeneracy with no need for application of the field. Moreover, the group velocities of the oppositely polarized modes become essentially different. The general reason is that the sublattice symmetry-breaking exchange at the uncompensated interface results in the symmetry-breaking between the two opposite-spin magnon modes. The dynamical polarization of the conduction electrons interacts with the oppositely polarized modes asymmetrically. This effect takes place both for AF/N and AF/S heterostructures. It has two contributions. The first one is from quasiparticles, which dominate at small magnon wave numbers $k$. The second contribution is caused by the interaction of magnons with a superconducting condensate and is explained by the appearance of the magnon-cooparon composite particles, analogously to the case of F/S hybrids\cite{Bobkova2022}. Further we show that damping of the magnon modes can also be renormalized by the back action of the metal, that is by magnon-assisted processes of the electronic quasiparticles spin flip. Finally, we demonstrate  the appearance of the additional modes in the magnon spectrum due to the hybridization between the magnons and electron paramagnetic resonance mode, which is caused by the exchange field induced in the metal by the antiferromagnet itself. 

The paper is organized as follows. In Sec.~\ref{model} the system under consideration and the model under study are described. In Sec.~\ref{formalism} we formulate the theoretical framework to treat the problem. In Sec.~\ref{results} the results for the magnon spectrum renormalization are presented and discussed: Sec.~\ref{results}A gives a general overview of the renormalization and splitting effects; Sec.~\ref{results}B discusses the physical reasons and temperature dependence of the renormalization; Sec.~\ref{results}C is devoted to the dependence of the renormalization effect on the value of the effective exchange field induced in the S(N) layer; in Sec.~\ref{results}D the dependence on the applied magnetic field is discussed.  A short summary of our research is given in Sec.~\ref{conclusions}.

\section{System and model}

\label{model}

\begin{figure}[h!]
 \centerline{$
 \begin{array}{c}
 \includegraphics[width=3.2in]{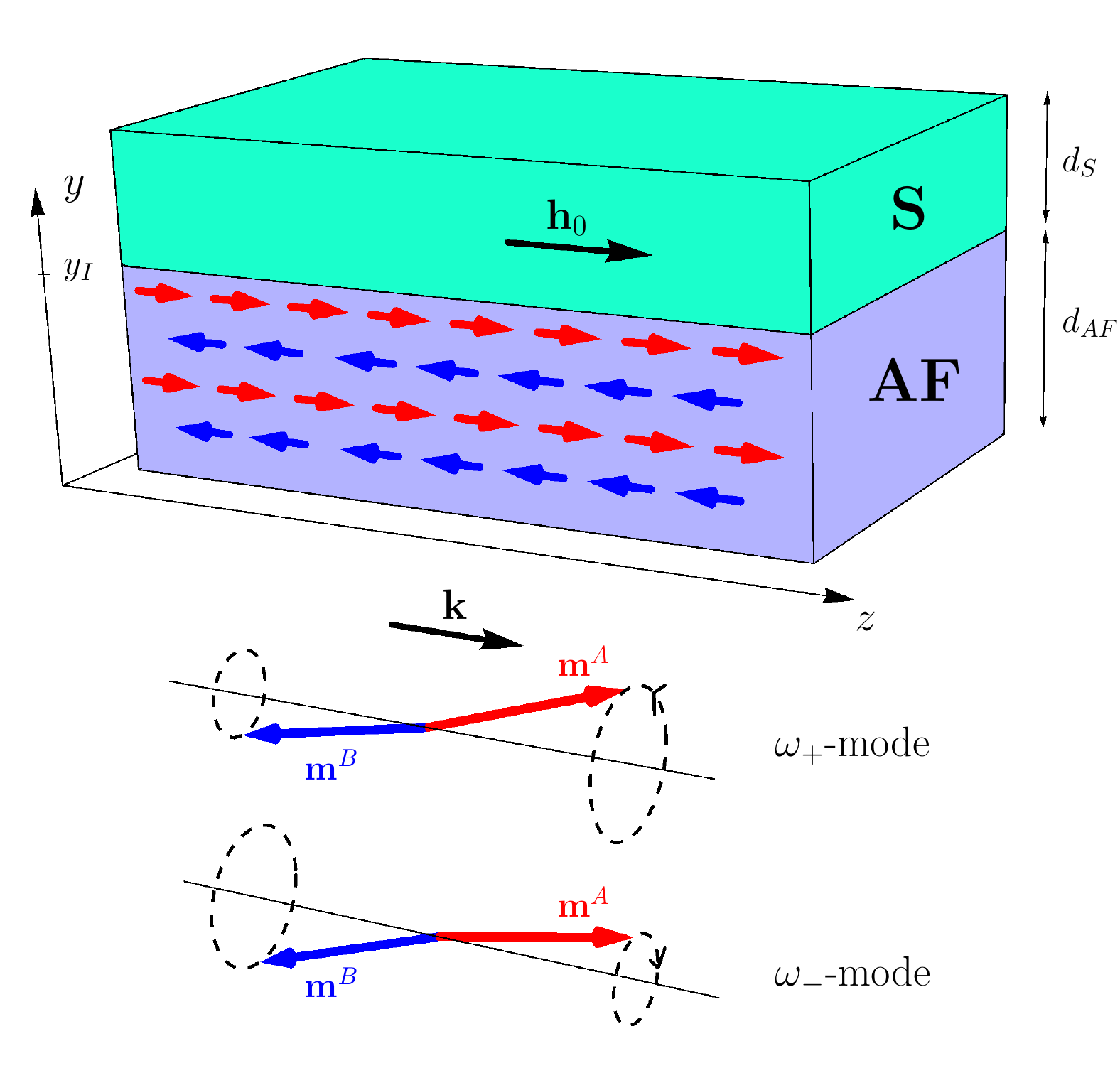} 
 \end{array}$}
 \caption{Sketch of the AF/S bilayer under consideration and a schematic illustration of the circularly-polarized magnon modes, which are energetically degenerate without a coupling to the S(N) layer and at zero applied magnetic field $\bm H=0$. The equilibrium (without a magnon) effective exchange field induced in the S layer by proximity to the AF insulator is denoted as $\bm h_0$.}
\label{sketch} 
\end{figure}

The model system that we consider is shown in Fig.~\ref{sketch}. It is a thin film bilayer consisting of an antiferromagnetic insulator with homogeneous N\'eel order interfaced to a conventional spin-singlet superconductor. The AF/N interface is considered as a limiting case, corresponding to temperatures above the superconducting critical temperature. We assume uncompensated magnetic moments at the AF/S interface, that is the interface possesses finite magnetization.  
Under these conditions, we expect the uncompensated AF to induce a spin-splitting field in the superconductor~\cite{Kamra2018} via an interfacial exchange interaction. It is assumed that the thickness of the S film $d_S$ is smaller than the superconducting coherence length $\xi_S$. Then the induced spin-splitting can be considered as uniform along the $y$-direction. In general the magnetic proximity effect at an AF/S uncompensated interface is not  reduced to the effective exchange only \cite{Kamra2018}, analogously to the case of the ferromagnetic insulator/superconductor interface \cite{Cottet2009,Eschrig2015}. However, in the framework of the present study we neglect other terms which can be viewed as additional magnetic impurities in the superconductor and focus on the effect of the exchange field. 

We treat S within the quasiclassical framework and solve the Eilenberger equation to obtain the Green's function and, consequently, the electron spin polarization, which is expressed in terms of the Green's function. The AF insulator is treated via the two-sublattice Landau-Lifshitz-Gilbert (LLG) description. The sublattices $A$ and $B$ are marked by the corresponding superscripts. The two subsystems are coupled due to the interfacial exchange, which is assumed nonzero between S and the sublattice A. The AF film is also assumed to be thin with uniform N\'eel order along the $y$-direction. The interface exchange leads to a spin-splitting term in the Eilenberger equation describing S and a spin torque term describing the magnetization dynamics for the AF sublattice A. The overall system dynamics is determined by solving the coupled equations self-consistently. 
 
We wish to examine wavevector-resolved excitations of the hybrid. Thus we can obtain the complete information needed for examining arbitrary wavepackets generated by a given experimental method. To this end, we assume the existence of a spin wave with wavevector $\mathbf k$ in the AF [Fig.~\ref{sketch}] such that the magnetization unit vector $\mathbf{m}^{A,B}(\mathbf{r},t) = \mathbf{m}_0^{A,B} + \delta \mathbf{m}^{A,B}(\mathbf{r},t)$ consists of the equilibrium part $\mathbf{m}_0^{A,B} = \pm \mathbf{e}_z$ and the excitation part $\delta \mathbf{m}_\pm^{\nu}(\mathbf{r},t) = \delta m_\pm^{\nu} \left[ \cos(\mathbf k \mathbf r + \omega t) \mathbf{e}_x \pm \sin(\mathbf k \mathbf r + \omega t) \mathbf{e}_y \right] \exp(- \kappa t)$, where $\nu = A,B$. Unlike the ferromagnetic case, in AFs two magnon modes\cite{Rezende_review} are possible for the same wavevector $\mathbf k$. Please note that in the last expression the subscript $\pm$ does not correspond to the sublattice $A(B)$ index, and denotes two different magnon modes.  They are described by the excitation parts $\delta \mathbf{m}_\pm^{A,B}(\mathbf{r},t)$ and are called  $\omega_\pm$ modes. The magnon modes are schematically shown in Fig.~\ref{sketch}. In these modes the two sublattice magnetizations
are nearly opposite to each other. They precess circularly, i.e., counterclockwise in one of the modes and clockwise in the other. In the quantum language they correspond to magnons carrying the opposite spins. In the absence of the applied field and proximity to the superconductor (or normal metal) these modes are degenerate. The effect of their own dipolar fields can also remove the degeneracy and strongly hybridizes the spin carried by the modes \cite{Kamra2017_2}. Here we neglect the effect of dipolar fields because according to our numerical estimates the coupling to the S(N) film generates larger splitting of the modes, of the order of tens of $\mathrm{GHz}$, while the splitting due to the dipolar interaction was estimated as $\sim \mathrm{GHz}$\cite{Kamra2017_2}. In addition the proximity-induced splitting manifests a very strong temperature dependence at low (superconducting) temperatures, which makes it possible to distinguish it.  We assume magnon wavevector $\mathbf k$ to be in the plane of the AF/S interface. 

\section{Formalism}

\label{formalism}

\subsection{Description of magnons in AF}

The magnetization dynamics is described within the Landau-Lifshitz-Gilbert framework.  The LLG equation can be written for each sublattice separately~\cite{Kamra2018_2}:
\begin{eqnarray}
\frac{\partial\bm m^i}{\partial t} = -\gamma \bm m^i \times \bm H_{\mathrm{eff}}^i + \sum \limits_j \alpha_{ij} \bm m^i \times \frac{\partial\bm m^j}{\partial t} + \bm N^i,~~~~~~
\label{LLG}
\end{eqnarray}
where $i=A,B$ is the sublattice index and $- \gamma$ with $\gamma>0$ is the AF gyromagnetic ratio,  $\bm{H}_{\mathrm{eff}}^i = Km_z^i \bm e_z + A \bm \nabla^2 \bm m^i - J_{\mathrm {AF}}\bm m^{\bar i}+\bm H$ is the effective magnetic field in the AF containing  an external magnetic field $\bm H$, easy-axis anisotropy with easy axis along the $z$-direction and anisotropy constant $K$, exchange stiffness $A$ and antiferromagnetic exchange coupling between two sublattices with coupling constant $J_{\mathrm {AF}}>0$. $\bar i=B(A)$ for $i=A(B)$. $\alpha_{ij}$ is the $2 \times 2$ Gilbert dissipation matrix\cite{Kamra2018_2,Yuan2019}, which can be characterized by two real positive numbers $\alpha$ and $\alpha_c$ as follows: $\alpha_{AA} = \alpha_{BB}=\alpha$ and $\alpha_{AB}=\alpha_{BA}=\alpha_c$.  The last term in Eq.~\eqref{LLG} represents the torque experienced by the sublattice. 

The torque $\bm N^i$ can be calculated starting from the effective exchange interaction between the spin densities on the two sides of the AF/S interface:
\begin{eqnarray}
H_{int} = - \int d^2 \bm r J \bm S^A \cdot \bm s,
\label{interface_ham}
\end{eqnarray}
where $\bm s$ is the electronic spin density operator in the S film, $\bm S^A$ is the localized spin operator in the AF film, belonging to the sublattice A. $J$ parameterizes interfacial exchange interaction between AF and S and the integration is performed over the 2D interface.  It has been shown \cite{Kamra2018} that this exchange interaction hamiltonian results in the appearance of the exchange field $\bm h(\bm r) = J M \bm m^A (\bm r)/(2\gamma d_S)$ in the S film, where $M$ is the sublattice saturation magnetization. 

From Eq.~(\ref{interface_ham}) one obtains the additional contribution to the Landau-Lifshitz-Gilbert equation written in the form of an interface torque acting on the magnetization:
\begin{eqnarray}
\bm N^A_{\mathrm{int}} = J \delta(y-y_I) \bm m^A (y) \times \bm s ,~~~ 
\bm N^B = 0,
\label{torque_general}
\end{eqnarray}
where the interface is located at $y=y_I$. We assume that the antiferromagnetic film is thin and its magnetization for a given sublattice is homogeneous in the $y$-direction. In this case Eqs.~(\ref{LLG}) and (\ref{torque_general}) can be averaged over the thickness $d_{AF}$ of the AF film. For the averaged torque we thus obtain
\begin{eqnarray}
\bm N^A = \frac{J \bm m^A \times  \bm s}{d_{\mathrm{AF}}}=\tilde J \bm m^A \times  \bm s,~~~ \bm N^B = 0.
\label{torque_averaged}
\end{eqnarray}
Here $\tilde{J} \equiv J/d_{\mathrm{AF}}$.  
We expand the electronic spin polarization over the three perpendicular vectors as
\begin{eqnarray}
\bm{s} = s_0 \bm e_z + \delta s_{\parallel} \delta \bm{m}^A + \delta s_{\perp} (\delta \bm{m}^A \times \bm e_z),
\label{electron_polarization}
\end{eqnarray}
where $s_0$ is the equilibrium value of the electron spin polarization in the superconductor, corresponding to the absence of the magnon. $\delta s_\parallel$ and $\delta s_\perp$ describe the dynamic corrections to the spin polarization due to the magnon. The components $s_0$, $\delta s_\parallel$ and $\delta s_\perp$ are in general functions of $(\omega, \bm k)$ and are calculated in the framework of the Usadel equations below. Substituting the expressions for $\bm{s}$ and $\bm{m}^{A,B}(\bm{r},t) = \pm \bm e_z + {\rm Re}[\delta \bm{m}^{A,B}e^{i\bm k \bm r +i \omega t}]$ in Eq.~\eqref{LLG} above and linearizing this equation with respect to $\delta \bm m^{A,B}$, we obtain
\begin{widetext}
\begin{eqnarray}
\left(
\begin{array}{cc}
-\omega - i \kappa \pm B_+ \pm \delta \omega_\pm \pm i \omega \alpha - i \chi_\pm & \pm \gamma J_{\mathrm{AF}} \pm i \omega \alpha_c \\
\mp \gamma J_{\mathrm{AF}} \mp i \omega \alpha_c & -\omega - i \kappa \mp B_- \mp i \omega \alpha 
\end{array}
\right)
\left(
\begin{array}{c}
\delta  m_\pm^A \\
\delta  m_\pm^B
\end{array}
\right) = 0,
\label{eq:modes}
\end{eqnarray}
\end{widetext}
where $B_\pm = \gamma(K+J_{\mathrm{AF}}+Ak^2 \pm H)=B\pm \gamma H$. Eq.~(\ref{eq:modes}) decouples giving two independent circularly polarized magnon modes $\delta  m_\pm^{A(B)} = \delta  m_x^{A(B)} \pm i\delta  m_y^{A(B)}$. Terms $\delta \omega_\pm$ and $\chi_\pm$ result from the torque acting on the AF from the superconductor. $\delta \omega_\pm = \tilde J[\delta s_\parallel(\pm \omega, \pm \bm k)-s_0(\pm \omega, \pm \bm k)]$ and $\chi_\pm = \tilde J \delta s_\perp(\pm \omega,\pm \bm k)$.
Neglecting the small terms resulting from $(\omega \alpha)^2$, $(\omega \alpha_c)^2$  and $\chi_\pm^2$ in Eq.~(\ref{eq:modes}) we obtain the following expression for the mode frequencies, which are determined by the real part $\omega$ of the solution $\omega_\pm + i \kappa_\pm$, providing zero determinant of the matrix in Eq.~(\ref{eq:modes}): 
\begin{eqnarray}
\omega_\pm = \pm \bigl( \frac{\delta \omega_\pm}{2} + \gamma H \bigr)+ 
\sqrt{\bigl( \frac{\delta \omega_\pm}{2} )^2+\omega_0^2  + \delta \omega_\pm B},
\label{eq:frequencies}
\end{eqnarray}
where $\omega_0 = \gamma\sqrt{(K+J_{\mathrm{AF}}+Ak^2)^2-J_{\mathrm{AF}}^2}$ is the AF magnon frequency at $H=0$ and without any contact with the S(N) layer. It is seen from Eq.~(\ref{eq:frequencies}) that the interaction of the AF magnons with polarized electrons in the adjacent metal layer removes degeneracy of the magnon modes even at zero magnetic field.  

The magnon decay rate $\kappa$ is also renormalized by the influence of the metal layer. The renormalized $\kappa$ is determined by the imaginary part $\kappa$ of the solution $\omega_\pm + i\kappa_\pm$ providing zero determinant of the matrix in Eq.~(\ref{eq:modes}) and up to the linear order with respect to $\alpha$, $\alpha_c$ and $\chi_\pm$ takes the form:
\begin{eqnarray}
\kappa_\pm = \frac{\alpha \omega \bigl(B + \frac{\delta \omega_\pm}{2} \bigr) - J_{\mathrm{AF}}\omega \alpha_c \mp \frac{\chi_\pm}{2}(B_- \pm \omega)}{\sqrt{\bigl( \frac{\delta \omega_\pm}{2} )^2+\omega_0^2  + \delta \omega_\pm B}}.
\label{eq:decay_rate}
\end{eqnarray}

\subsection{Electron spin polarization in S(N)}

Now our goal is to calculate the electron spin polarization $\bm s$ in the superconductor. The approach closely follows Ref.~\onlinecite{Bobkova2022}, where it has been used for treating S/F bilayers. Here we provide the corresponding equations for consistency and generalize the calculations for the case of two magnon modes.  $\bm s$ can be calculated via the quasiclassical Green's function as
\begin{eqnarray}
\bm s = -\frac{N_F}{16} \int d\varepsilon {\rm Tr}_4 \Bigl[ \bm \sigma \tau_z \check g^K \Bigr],
\label{s}
\end{eqnarray}
where $\check g^K$ is the Keldysh component of the quasiclassical Green's function $\check g$, which is a $8 \times 8$ matrix in the direct product of Keldysh, spin and particle-hole spaces. It obeys the Usadel equation:
\begin{eqnarray}
iD \bm \nabla \bigl( \check g \otimes \bm \nabla \check g \bigr) = \bigl[ \varepsilon \tau_z - \bm h \bm \sigma \tau_z + \Delta i\tau_y, \check g \bigr]_\otimes,
\label{usadel_1}
\end{eqnarray}
where $[A,B]_\otimes = A\otimes B -B \otimes A$ and we work in the mixed $(\varepsilon,t)$ representation with $A \otimes B = \exp[(i/2)(\partial_{\varepsilon_1} \partial_{t_2} -\partial_{\varepsilon_2} \partial_{t_1} )]A(\varepsilon_1,t_1)B(\varepsilon_2,t_2)|_{\varepsilon_1=\varepsilon_2=\varepsilon;t_1=t_2=t}$. In case $ A[B](\varepsilon, t) =  A[B] (\varepsilon) \exp[i \Omega_{A[B]} t]$ the $\otimes$-product is reduced to $ A(\varepsilon,t) \otimes B(\varepsilon, t) = A (\varepsilon-\Omega_B/2)  B(\varepsilon+\Omega_A/2,t)\exp[i (\Omega_A + \Omega_B) t]$. $\tau_{x,y,z}$ and $\sigma_{x,y,z}$ are Pauli matrices in particle-hole and spin spaces, respectively. $\Delta $ is the superconducting order parameter. The explicit structure of the Green's function in the Keldysh space takes the form:
\begin{eqnarray}
\check g = 
\left(
\begin{array}{cc}
\check g^R & \check g^K \\
0 & \check g^A
\end{array}
\right),
\label{keldysh_structure}
\end{eqnarray}
where $\check g^{R(A)}$ is the retarded (advanced) component of the Green's function and $g^K$ is the Keldysh component. Further we express the Keldysh part of the Green's function  via the retarded, advanced Green's functions and the distribution function $\check \varphi$ as follows: $\check g^K = \check g^R \otimes \check \varphi - \check \varphi \otimes \check g^A$.

The exchange field is taken in the form of a time-independent component and a circularly polarized magnon:
\begin{eqnarray}
\bm h_\pm = h_0 \bm e_z + \delta h \cos(\bm k \bm r + \omega t) \bm e_x \pm \delta h \sin(\bm k \bm r + \omega t) \bm e_y ~~
\label{h}
\end{eqnarray}
Then
\begin{eqnarray}
\delta \bm  h_\pm \bm \sigma = \delta h_\pm e^{\mp i(\bm k \bm r + \omega t)\sigma_z}\sigma_x .
\label{h2}
\end{eqnarray}
Please note that $\bm h_\pm$, respectively induced by the $\omega_\pm$ magnon modes, does not have a sublattice index.  In the framework of the considered simplified model the superconductor is only coupled to the $A$-sublattice of the antiferromagnet and the induced exchange field is just an imprint of this sublattice $\bm h_\pm = JM \bm m_\pm^A/(2\gamma d_S)$. In reality, of course a weaker exchange coupling of the exponentially decaying electron wave function to the $B$-sublattice is also possible. However, it does not change the physics qualitatively, and only reduces to some extent the amplitude of the effective exchange field $\delta h_{\pm} \propto(\delta m_\pm^A - w \delta m_\pm^B) $, where $w$ is a factor taking into account the relative density of the electron wave function at the $A$ and $B$ sublattices. The quasiclassical Green's function is to be found in the form: $\check g_\pm = \check g_0 + \delta \check g_\pm$, where $\check g_0$ is the Green's function in the absence of the magnon and $\delta g_\pm$ is the first order correction with respect to $\delta \bm h_\pm$. Taking into account that $\bm \nabla g_0 = 0$ (we assume that in the absence of the magnon the bilayer is spatially homogeneous along the interface) from Eq.~(\ref{usadel_1}) we obtain the following equation for $\delta \check g_+$:
\begin{eqnarray}
i D \check g_0 \otimes \nabla^2 \delta \check g_+ = \bigl[  \varepsilon \tau_z - h_0 \bm e_z \sigma_z \tau_z + \tau_z \hat \Delta, \delta \check g_+ \bigr]_\otimes - \nonumber \\
\bigl[  \delta h e^{-i(\bm k \bm r + \omega t)\sigma_z} \sigma_x \tau_z, \check g_0 \bigr]_\otimes ~~~~~~~~
\label{delta_g1}
\end{eqnarray}
Introducing the unitary operator $\hat U = e^{-i(\bm k \bm r + \omega t)\sigma_z/2}$ we can transform the  Green's function as follows:
\begin{eqnarray}
\delta \check g_+ = U \otimes \delta \check g_m \otimes U^{\dagger} .
\label{transform}
\end{eqnarray}
In the case where the system is spatially homogeneous except for the magnon, $\delta \check g_m$ does not depend on coordinates. Then 
\begin{eqnarray}
\nabla^2 \delta \check g_+ = -\frac{k^2}{2} \check U \otimes \bigl( \delta \check g_m - \sigma_z \delta \check g_m \sigma_z  \bigr) \otimes U^\dagger = -k^2 \delta \check g_+,~~~~
\label{delta_g2}
\end{eqnarray}
where when passing to the second equality it is used that $\delta \check g_+ = \delta \hat g_x \sigma_x + \delta \hat g_y \sigma_y$ and has no $z$-component in the spin space according to the spin structure of the magnon exchange field $\delta \bm h_+$.

From the normalization condition $\check g \otimes \check g = 1$ it follows that $\check g_0 \otimes \delta \check g_+ = - \delta \check g_+ \otimes \check g_0$. It gives us $\check g_0 \otimes \delta \check g_+ = (1/2)[\check g_0, \delta \check g_+]_\otimes$. Eq.~(\ref{delta_g1}) takes the form:
\begin{eqnarray}
\bigl[  \varepsilon \tau_z + i \frac{D k^2}{2} \check g_0 - h_0 \bm e_z \sigma_z \tau_z + \tau_z \hat \Delta, \delta \check g_+ \bigr]_\otimes - \nonumber \\
\bigl[  \delta h e^{-i(\bm k \bm r + \omega t)\sigma_z} \sigma_x \tau_z, \check g_0 \bigr]_\otimes = 0. ~~~~~~~~
\label{delta_g3}
\end{eqnarray}
From Eq.~(\ref{delta_g3}) the following equation for  $\delta \check g_m$ is obtained:
\begin{eqnarray}
\bigl[ \hat \Lambda_d \tau_z + \hat \Lambda_{od} i\tau_y, \delta \check g_m \bigr] = \delta h \bigl[ \sigma_x \tau_z, \nonumber \\
 \frac{1}{2}\bigl((g_{0,s}+g_{0,a}\sigma_z)\tau_z + (f_{0,s}+f_{0,a}\sigma_z)i\tau_y \bigr) \bigr].
\label{delta_g4}
\end{eqnarray}
It does not contain time dependence and $\otimes$-products. In Eq.~(\ref{delta_g4}) we use the following definitions
\begin{eqnarray}
g_{0,s(a)} = g_{0,\uparrow}(\varepsilon+\frac{\omega}{2}) \pm g_{0,\downarrow}(\varepsilon-\frac{\omega}{2}), \label{delta_g_def1} \\
f_{0,s(a)} = f_{0,\uparrow}(\varepsilon+\frac{\omega}{2}) \pm f_{0,\downarrow}(\varepsilon-\frac{\omega}{2}), 
\label{delta_g_def2}
\end{eqnarray}
where $g_{0,\uparrow(\downarrow)}$ represent the bulk Green's functions for the superconductor in the exchange field $h_0$: 
\begin{eqnarray}
g_{0,\uparrow(\downarrow)}^R = \frac{|\varepsilon \mp h_0|}{\sqrt{(\varepsilon + i \delta \mp h_0)^2-\Delta^2}} \label{delta_g_def3} \\
f_{0,\uparrow(\downarrow)}^R = \frac{\Delta {\rm sgn}(\varepsilon \mp h_0)}{\sqrt{(\varepsilon + i \delta \mp h_0)^2-\Delta^2}},
\label{delta_g_def4}
\end{eqnarray}
and $g(f)_{0,\uparrow(\downarrow)}^A = -g(f)_{0,\uparrow(\downarrow)}^{R*}$, $g(f)_{0,\uparrow(\downarrow)}^K = [g(f)_{0,\uparrow(\downarrow)}^R - g(f)_{0,\uparrow(\downarrow)}^A] \tanh[\varepsilon/2T]$.
\begin{eqnarray}
\hat \Lambda_d = \Lambda_{d}^0 + \Lambda_{d}^z \sigma_z = \nonumber \\
\varepsilon + \frac{i D k^2}{4}g_{0,s} + \bigl( \frac{\omega}{2}-h_0 + \frac{i D k^2}{4}g_{0,a} \bigr)\sigma_z,
\label{delta_g_def5} \\
\hat \Lambda_{od} = \Lambda_{od}^0 + \Lambda_{od}^z \sigma_z = \nonumber \\
\Delta + \frac{i D k^2}{4}f_{0,s} + \frac{i D k^2}{4}f_{0,a} \sigma_z .
\label{delta_g_def6}
\end{eqnarray}
Solving Eq.~(\ref{delta_g4}) we obtain:
\begin{eqnarray}
\delta \check g_m = \delta g_{mx} \sigma_x \tau_z + \delta f_{mx} \sigma_x i\tau_y , 
\label{delta_g_sol1}
\end{eqnarray}
where
\begin{eqnarray}
\delta g_{mx} = \frac{\delta h \bigl[ f_{0,s}\Lambda_{od}^z - g_{0,a}\Lambda_d^0 \bigr]}{2 \bigl[ \Lambda_d^0\Lambda_d^z - \Lambda_{od}^0\Lambda_{od}^z \bigr]}
\label{delta_g_sol2}
\end{eqnarray}
The distribution function also acquires a correction due to the magnon: $\check \varphi_+ = \check \varphi_0 + \delta \check \varphi_+$. It is convenient to work with the transformed distribution function $\hat U^\dagger \check \varphi_+ \hat U = \check \varphi_m + \delta \varphi_m$, which does not depend on time and spatial coordinates. Here $\check \varphi_m = (1/2)[\varphi_{m,s}+\varphi_{m,a}\sigma_z]$ with $\varphi_{m, s(a)} = \tanh[(\varepsilon+\omega/2)/2T] \pm \tanh[(\varepsilon-\omega/2)/2T]$  the result of the unitary transformation of the equilibrium distribution function $\check \varphi_0 = \tanh [\varepsilon/2T]$. From the Keldysh part of the Usadel equation (\ref{delta_g4}) we can obtain the following equation for the first order correction to the distribution function $\delta \check \varphi_m $:
\begin{eqnarray}
i Dk^2 [\delta \check \varphi_m - \check g_{m0}^R \delta \check \varphi_m \check g_{m0}^A] + \nonumber \\
\check g_{m0}^R [\check K, \delta \check \varphi_m] - [\check K, \delta \check \varphi_m] \check g_{m0}^A  + \nonumber \\
\check g_{m0}^R [\check \varphi_m, \delta h \sigma_x \tau_z] - [\check \varphi_m, \delta h \sigma_x \tau_z] \check g_{m0}^A = 0, 
\label{distribution_correction}
\end{eqnarray}
where 
\begin{eqnarray}
\check K = (\varepsilon+(\omega/2-h_0)\sigma_z )\tau_z + \Delta i \tau_y,
\label{distribution_def_1}
\end{eqnarray}
\begin{eqnarray}
\check g_{m0}^{R,A} = \hat U^\dagger \otimes \check g_0^{R,A} \otimes \hat U = \nonumber \\
(1/2)[(g_{0,s}^{R,A}+g_{0,a}^{R,A}\sigma_z)\tau_z + (f_{0,s}^{R,A}+f_{0,a}^{R,A}\sigma_z)i\tau_y],
\label{distribution_def_2}
\end{eqnarray}
The structure of Eq.~(\ref{distribution_correction}) dictates that 
\begin{eqnarray}
\delta \check \varphi_m = 
\left(
\begin{array}{cc}
0 & \delta \varphi_m^\uparrow \\
\delta \varphi_m^\downarrow & 0
\end{array}
\right).
\label{distribution_correction_2}
\end{eqnarray}
Substituting Eq.~(\ref{distribution_correction_2}) into Eq.~(\ref{distribution_correction}) we obtain the following result:
\begin{eqnarray}
\delta \varphi_m^\sigma = -\delta h \varphi_{m,a} \times ~~~~~~~~~ \nonumber \\
\frac{2h_{0\omega}G_{-,\sigma}+i \sigma D k^2 (g_{m\sigma}^R - g_{m \bar \sigma}^A)}{4h_{0\omega}^2 G_{-,\sigma} + 4 h_{0\omega} i \sigma D k^2 (g_{m\sigma}^R - g_{m \bar \sigma}^A) - (Dk^2)^2 G_{+,\sigma}},~~~~
\label{distribution_correction_3}
\end{eqnarray}
where we introduce the spin subband index $\sigma = \uparrow(\downarrow)$ in the subscripts/superscripts and $\sigma = \pm 1$ for spin-up(down) subbands, respectively, if it is as a factor. Here, $\bar \sigma = -\sigma$, $h_{0\omega} = \omega/2-h_0$, ${g(f)}_{m\sigma}^{R,A} = g(f)_{0,\sigma}^{R,A}(\varepsilon + \sigma \omega/2)$ and $G_{\pm,\sigma} = 1-g_{m\sigma}^R g_{m \bar \sigma}^A \pm f_{m\sigma}^R f_{m \bar \sigma}^A$.

The electron spin polarization $\bm s_+$, induced in the superconductor by the equilibrium magnetization $\bm m_0^A$ and the magnon mode $\delta \bm m_+^A$ is calculated from Eq.~(\ref{s}). It can be decomposed into three orthogonal components according to Eq.~(\ref{electron_polarization}). The corresponding components take the form:
\begin{eqnarray}
s_0  = -\frac{N_F}{4} \int \limits_{-\infty}^\infty d \varepsilon \tanh \frac{\varepsilon}{2T} {\rm Re}\bigl[ g_{0,\uparrow}^R - g_{0,\downarrow}^R \bigr],
\label{spin_0}
\end{eqnarray}
\begin{eqnarray}
\delta s_\parallel(\omega, \bm k)  = -\frac{N_F h_0}{8\delta h} \int \limits_{-\infty}^\infty d \varepsilon \Bigl\{ 2 \varphi_{m,s} {\rm Re} [\delta g_{mx}^R] + \nonumber \\
 \sum \limits_\sigma (g_{m\sigma}^R - g_{m \bar \sigma}^A) \delta \varphi_m^\sigma  \Bigr\},
\label{spin_parallel}
\end{eqnarray}
\begin{eqnarray}
\delta s_\perp(\omega, \bm k)  = \frac{N_F h_0}{8\delta h} \int \limits_{-\infty}^\infty d \varepsilon \Bigl\{ 2 \varphi_{m,a} {\rm Im} [\delta g_{mx}^R] + \nonumber \\
i \sum \limits_\sigma \sigma (g_{m\sigma}^R - g_{m \bar \sigma}^A) \delta \varphi_m^\sigma\Bigr\}.
\label{spin_perpendicular}
\end{eqnarray}
The electron spin polarization induced by the mode $\delta \bm m^A_-$ can be obtained from Eqs.~(\ref{spin_parallel}) and (\ref{spin_perpendicular}) by the substitution $\omega \to -\omega$ and $\bm k \to -\bm k$. 

\section{Results and discussion}

\label{results}

\subsection{Splitting and renormalization of the AF magnon mode}

\begin{figure}[h!]
 \centerline{$
 \begin{array}{c}
 \includegraphics[width=3.6in]{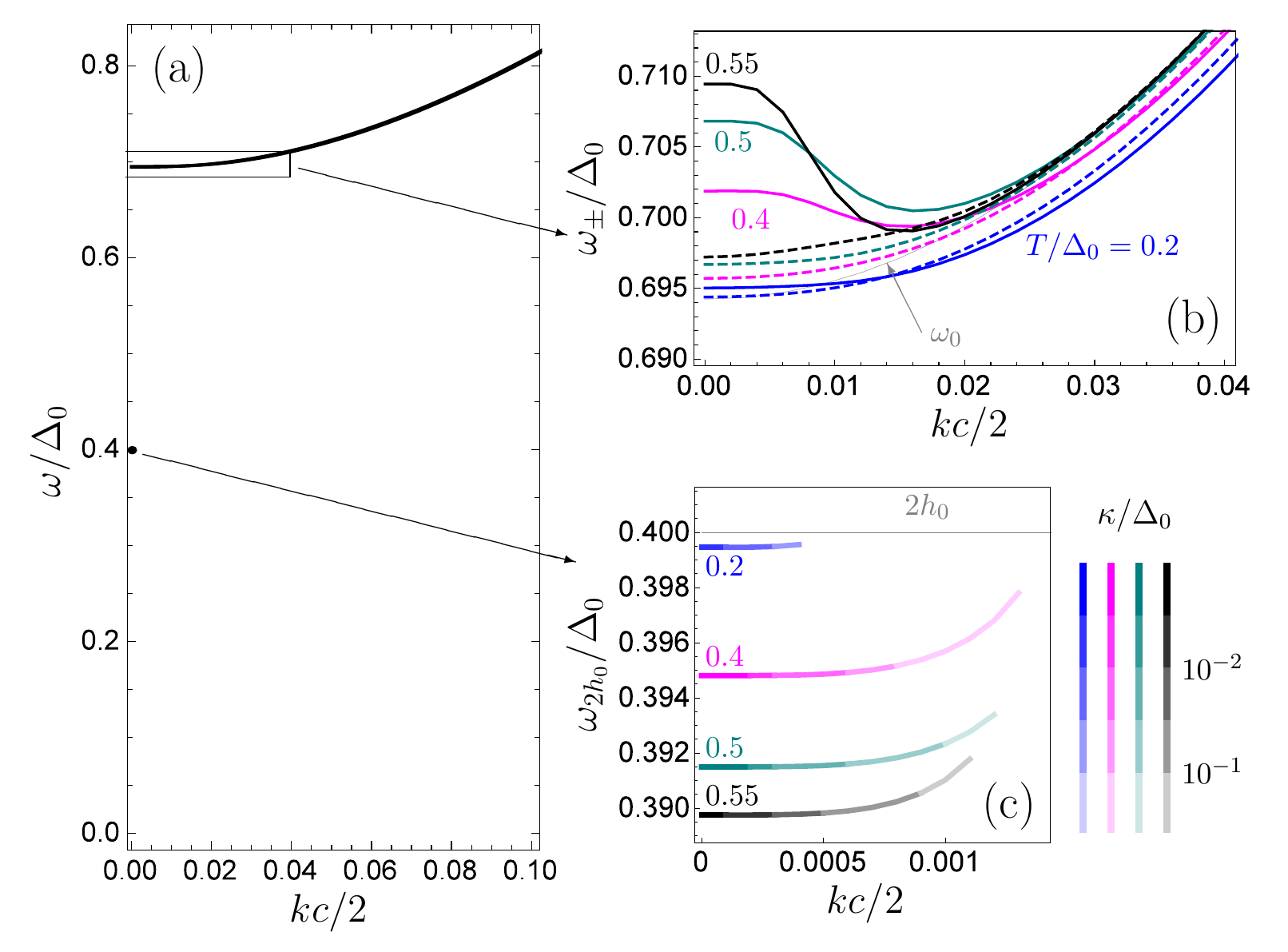} 
 \end{array}$}
 \caption{Magnon dispersion $\omega(k)$ for $h_0=0.2\Delta_0$. $\omega$ is measured in units of $\Delta_0$ and $k$ is measured in units of $2c^{-1}$, where $c$ is a length scale of the order of the interatomic distance. We take $c=3.3 \mathrm{\AA}$ in accordance with the lattice period along the $\hat c$-axis of $\mathrm{MnF_2}$ \cite{Griffel1950}. Panel (a) represents the general view of the magnon modes dispersion in the AF/S bilayer at $T=0.2 \Delta_0$. Panel (b) represents the temperature evolution of the dispersion in the rectangular region in panel (a) on a large scale. Different colors correspond to different temperatures. The $\omega_+$-mode is plotted by solid lines and the $\omega_-$-mode is plotted by  dashed lines. The original AF dispersion $\omega_0(k)$ is shown by grey line in panel (b). Panel (c) is the zoom of the $\omega_{2h_0}$-mode, which is seen as a black point in panel (a). Different colors are again different temperatures. The decay rate is encoded by the brightness of the corresponding line. We take $\beta = 0.05 h_0$, $\alpha=2\times 10^{-4}$, $\alpha_c=10^{-4}$, $\xi_S=10nm$, $\Delta_0=18K$, $\gamma K=0.061\Delta_0$, $\gamma J_{AF}=3.92\Delta_0$, $\gamma A(c/2\xi_S)^2=2.26\Delta_0$ throughout the paper.}
\label{fig:dispersions_1} 
\end{figure}

The dispersion relations for the both AF magnon modes $\omega_\pm(\bm k)$ at $H=0$ are presented in Fig.~\ref{fig:dispersions_1}. Panel (a) represents the general view of the mode frequencies on a large scale of the order of $\Delta_0$, where $\Delta_0$ is a value of the superconducting gap in the bulk superconductor at $T=0$ and without a proximity to the antiferromagnet. The numerical parameters of the AF/S bilayer are taken to be close to the material parameters of $\mathrm{MnF_2}$ (AF) \cite{Rezende_review}  and $\mathrm{Nb}$ (S). The upper mode is the original AF mode, split and renormalized by superconductivity. While the characteristic AF frequencies $\omega_0 \sim \Delta_0$ are of the order of $\mathrm{THz}$, all the effects of splitting and renormalization are caused by  $\delta \omega_\pm$, as is seen from Eq.~(\ref{eq:modes}). $\delta \omega_\pm \propto \tilde J \delta s_\parallel$. The characteristic value of $\delta s_\parallel$ in the superconducting state is of the order of $N_F \Delta_0$ (in the normal state it is $N_F h_0$). Therefore, the quantity $\delta \omega_\pm \propto \tilde J N_F \Delta_0 \equiv \beta$. Therefore, the parameter $\beta$ quantifies the back action of the electron polarization induced in the superconductor by the magnon on the magnon itself.  Taking $d_{\mathrm{AF}} = d_S $, $M \sim 230$G, $\gamma =2 \mu_B/\hbar$  and $N_F \approx 1.3 \times 10^{35} \mathrm{erg^{-1}cm^{-3}}$ for $\mathrm{Nb}$, we obtain for chosen materials $\beta \sim 0.05 h_0$. The equilibrium effective exchange field in the S(N) layer for the results presented in Fig.~\ref{fig:dispersions_1} is assumed to be $h_0=0.2\Delta_0$. This value corresponds to the available estimates of the effective field at the uncompensated AF/N interfaces of real materials\cite{Kamra2018}. 

For the chosen parameters we obtain that the superconductivity-induced splitting of the AF magnon mode is of the order of tens of $\mathrm{GHz}$. For this reason panel (b) represents the upper mode on the appropriate scale. First of all, it is seen from Fig.~\ref{fig:dispersions_1}(a) that the magnon modes are split even at zero magnetic field. Both $\omega_\pm(\bm k=0)$ and the magnon velocity $\bm v_{m,\pm} = d\omega_\pm/d \bm k$ are renormalized by the proximity to the adjacent metal. The physical reason for this splitting is that only the magnetization of $A$-sublattice $\bm m^A$ is coupled to the conductor. The amplitudes of magnon magnetizations at $A$-sublattice are different for $\bm m_\pm$ modes, see Fig.~\ref{sketch}. This results in different effective exchange fields, induced in the conductor by these modes, which, in its turn, results in nonequivalent back action of the electron polarization on the magnon.  

Moreover, due to the resonance shape of $\delta \omega_\pm(\omega)$ [see Eq.(\ref{omega_shift_normal}) below], the nonlinear equation (\ref{eq:frequencies}) gives additional solutions for magnon frequencies, which do not occur in the absence of the adjacent conductor. The corresponding magnon frequencies are located at $\omega_{2h_0} \approx 2h_0$. Their physical origin is from the hybridization of the magnon modes and the electron paramagnetic resonance mode in the conductor. However, the important point is that in the considered case the electron paramagnetic resonance mode does not occur in the S(N) layer by itself and appears due to the interaction with the antiferromagnet. They are represented in panel (c) of Fig.~\ref{fig:dispersions_1}. We discuss the additional modes in more details below.

\subsection{Physical reasons of the renormalization and temperature dependence}

\begin{figure}[h!]
 \centerline{$
 \begin{array}{c}
 \includegraphics[width=3.2in]{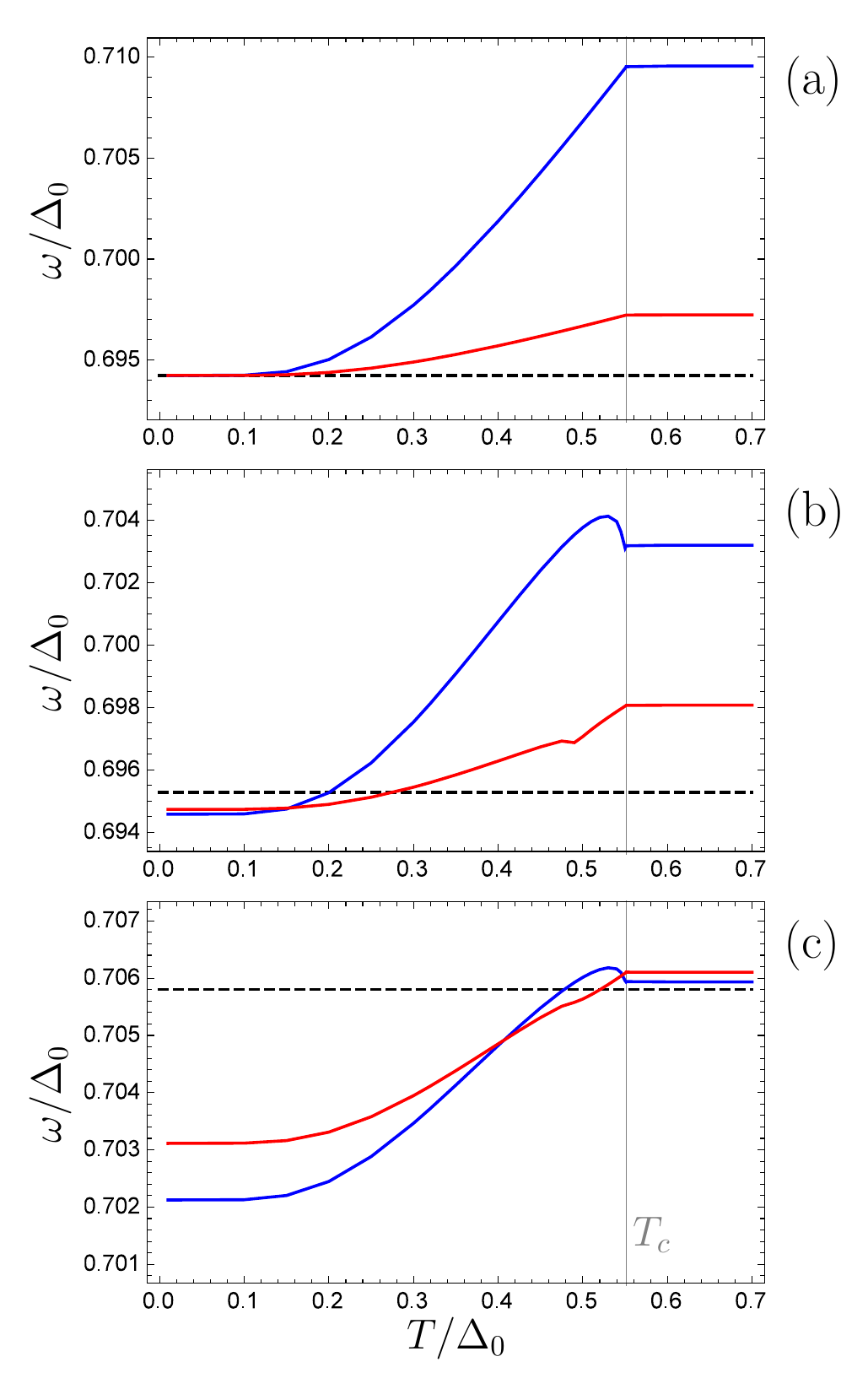} 
 \end{array}$}
 \caption{Dependence of the mode frequencies $\omega_+$(blue) and $\omega_-$(red) on temperature for three representative values of the magnon wave number $k$: (a)$k=0$, (b)$kc/2=0.009$, (c)$kc/2=0.03$. For all panels the black dashed line is $\omega_0$. The vertical grey line marks the superconducting critical temperature of the AF/S bilayer. $h_0=0.2\Delta_0$. }
\label{fig:k_T} 
\end{figure}

Let us first concentrate on the physical reasons of renormalization of the AF magnon mode. In panel (b) of Fig.~\ref{fig:dispersions_1} the corresponding frequencies are plotted for several temperatures. Different colors correspond to different temperatures. Solid and dashed lines of the same color represent $\omega_\pm$ for the same temperature. The renormalization of the magnon modes is originated from two different physical sources. One of them is the influence of quasiparticle polarization and the other one is the effect of an accompanying cloud of Cooper pairs. To study both contributions in more detail, in Fig.~\ref{fig:k_T} we plot the dependence of the mode frequencies $\omega_\pm$ on temperature for three representative values of the magnon wave number $k$.

From panel (a) of Fig.~\ref{fig:k_T} one can see that at $k=0$ (Kittel mode) the renormalization is fully suppressed at $T=0$. That indicates the quasiparticle nature of the effect. Due to the presence of the superconducting gap the quasiparticles in the S layer are suppressed at $T \to 0$. Physically the  renormalization of the $k=0$ AF frequency can be understood as follows. The AF magnetization, including the equilibrium magnetization and a magnon, induces an exchange field $\bm h$ in the superconductor. This exchange field polarizes quasiparticles. If the AF magnetization were static, the quasiparticle polarization would be aligned with it and would not generate a torque on the AF magnetization. However, due to the dynamical nature of the AF magnetization with finite frequency $\omega_0$ the instantaneous quasiparticle polarization in S is not aligned with it and works as an additional contribution to the anisotropy field, but different for both AF magnon modes.  The perpendicular component of the polarization is determined by the difference $\delta s_\parallel(\pm \omega, k=0)-s_0(\pm \omega, k=0)$, which disappears at  $ \omega_\pm/\Delta_0 \to 0$. This clearly indicates that the asymmetric shift of the $k=0$ magnons originates from the retardation effects. As  seen from Fig.~\ref{fig:k_T}(a) the $k=0$ frequency shift is the most pronounced in the normal state when the superconducting gap disappears. In this limit the induced electron polarization in N can be found analytically: 
\begin{eqnarray}
\delta s_{\parallel,N} = N_F \frac{\omega h_0 (\omega - 2 h_0)}{(\omega - 2h_0)^2+(Dk^2)^2}, \label{polarization_normal_1} \\
\delta s_{\perp,N} = -N_F \frac{\omega h_0 Dk^2}{(\omega - 2h_0)^2+(Dk^2)^2},
\label{polarization_normal_2}
\end{eqnarray}
while $s_{0,N} = 0$. For the $\omega_+$ mode this results in the following expressions for $\delta \omega$ and $\chi$:
\begin{eqnarray}
\delta \omega_{+,N} = N_F \tilde J \frac{\omega h_0 (\omega - 2 h_0)}{(\omega - 2h_0)^2+(Dk^2)^2}, \label{omega_shift_normal} \\
\chi_{+,N} = -N_F \tilde J \frac{\omega h_0 Dk^2}{(\omega - 2h_0)^2+(Dk^2)^2}.
\label{decay_shift_normal}
\end{eqnarray}
$\delta \omega_{-,N}(\omega) = \delta \omega_{+,N}(-\omega)$, and the same is valid for $\chi_{-,N}$. It is seen that $\delta \omega_+ \neq \delta \omega_-$. The most important contribution of the coupling $\delta \omega_\pm$ to the AF mode splitting in Eq.~(\ref{eq:frequencies}) is via the last term $\delta \omega_\pm B$. Disregarding the smaller terms containing $\delta \omega_\pm$ in this equation, at $H=0$ and $k=0$ we obtain that $\omega_\pm \approx \sqrt{\omega_0^2 + \delta \omega_\pm B} \approx  \sqrt{2 \gamma J_{\mathrm {AF}}(\gamma K+\delta \omega_\pm/2)}$. That is, $\delta \omega_\pm$ acts on the AF magnetization at $k=0$ in the same way as an anisotropy field does, but the contribution to the anisotropy field is different for  both magnon modes.

The similar shift of the Kittel mode $k=0$ also takes place for S/F bilayers \cite{Silaev2020,Bobkova2022}, but in that case the characteristic magnon frequencies are of the order of $\mathrm{GHz}$ and, consequently, $\omega_0/\Delta_0 \ll 1$. This leads to the fact that the effective dynamical contribution to the anisotropy field is small. Therefore, the $k=0$ shift is negligible for ferromagnetic magnons and the main renormalization there is due to the Cooper pairs \cite{Bobkova2022}. Here, for the AF problem, both sources of the renormalization are essential. For small magnon wave numbers $k$ the main contribution is due to quasiparticles. This contribution is suppressed at $k \sim \sqrt{|\omega-2h_0|/D}$, as  seen from Eq.~(\ref{polarization_normal_1}). 

To effectively generate equal-spin pairs screening the magnon, the spatial inhomogeneity of the exchange field in the superconductor is required \cite{Bobkova2022}. Consequently, this process is more efficient at large $k$, comparable to the inverse superconducting coherence length $\xi_S^{-1} = \sqrt{D/\Delta_0}$, which in our case corresponds to $k_{S}c/2 = \xi_S^{-1}c/2 = 0.0165$. Panel (c) of Fig.~{\ref{fig:k_T}} demonstrates the temperature dependence of the magnon frequencies renormalization at large $k>k_S$. It is seen that the temperature dependence is opposite to the $k=0$ limit, presented in panel (a). This is because the quasiparticle renormalization is already not effective at such wave numbers and the main effect is due to the screening of the magnons by the Cooper pairs. Panel (b) shows an intermediate case when the pair contribution only starts to come into play. 

\subsection{Additional modes and dependence of the renormalization on the value of the effective exchange}

\begin{figure}[h!]
 \centerline{$
 \begin{array}{c}
 \includegraphics[width=3.2in]{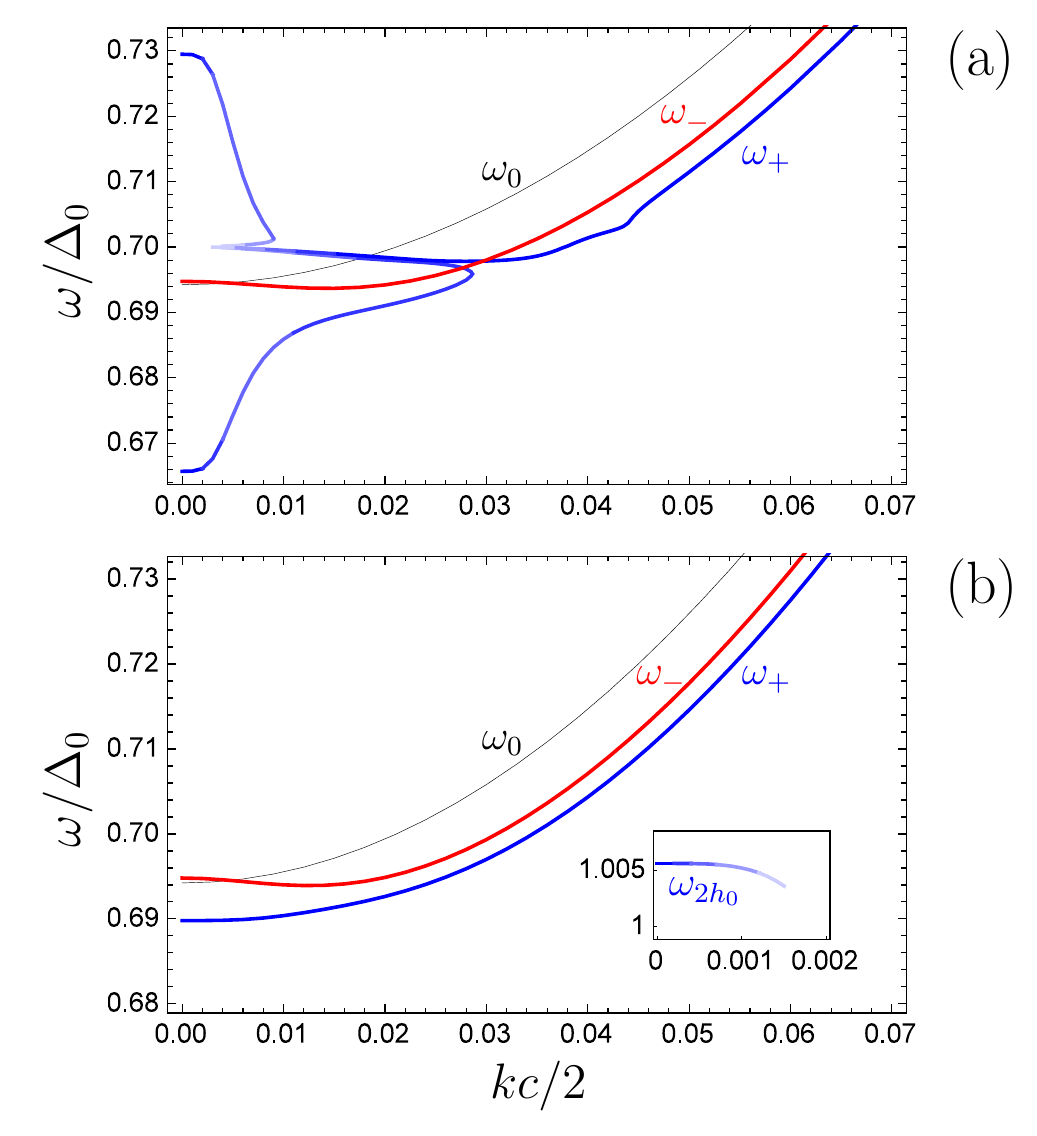} 
 \end{array}$}
 \caption{Magnon dispersion in the AF/S bilayer at (a) $h_0=0.35\Delta_0$ and (b) $h_0=0.5\Delta_0$. The inset to panel (b) demonstrates the $\omega_{2h_0}$ mode. Temperature $T=0.2\Delta_0$ for both panels. The decay rate $\kappa/\Delta_0$ is encoded by the brightness of the corresponding line, and the scale is the same as in Fig.~\ref{fig:dispersions_1}.}
\label{fig:h_cases} 
\end{figure}

Fig.~\ref{fig:h_cases} demonstrates how the renormalization of the magnons depends on the value of the induced exchange field $h_0$. While our basic consideration presented above corresponds to $2h_0 = 0.4 \Delta_0 < \omega_0(k=0)$, panels (a) and (b) of Fig.~\ref{fig:h_cases} illustrate two other possible cases $2 h_0 \approx \omega_0$ and $2h_0 > \omega_0$, respectively. It is worth mentioning that the amplitude of the effective exchange is $\sim d_S^{-1}$ and, therefore, can be tuned by varying the metal layer thickness. 

The induced exchange field $h_0$ gives rise to the electron paramagnetic resonance mode in the S(N) layer. In the considered case only spin up magnons with frequency $2h_0$ can be emitted and absorbed by the electron paramagnetic resonance mode. Spin up magnons correspond to the $\omega_+$-mode. For this reason only the $\omega_+$-mode can interact with the electron paramagnetic resonance mode.  It is seen from Eq.~(\ref{omega_shift_normal}) that the correction $\delta \omega_+$ has a resonance shape peaked at $2h_0$, which is originated from the interaction with the electron paramagnetic resonance in the N layer. On the contrary, the $\omega_-$-mode does not exhibit the resonant behavior at $\omega \approx 2h_0$. For the superconducting system the expression for $\delta \omega_+$ is more complicated, but it also has a similar resonance shape.  That results in additional solutions of nonlinear Eq.~(\ref{eq:frequencies}) located at $\omega_{2h_0} \approx 2h_0$. Moreover, if $h_0$ is chosen in such a way that $\omega_0(k=0)$ is close to $\omega_{2h_0}$, the $\omega_+$-mode interacts strongly with the electron paramagnetic resonance mode giving rise to the peculiar behavior shown in Fig.\ref{fig:h_cases}a. The details of the corresponding solution are presented in the Appendix. 

When $\omega_0(k=0)$ and $\omega_{2h_0}$ are well separated, the corresponding additional branches are above or below the main AF split modes and do not disturb them, as  shown in Fig.~\ref{fig:dispersions_1}(a) and Fig.~\ref{fig:h_cases}(b). In principle, the additional mode, originated from the electron paramagnetic resonance can serve as an experimental probe of the effective exchange field induced in the metal layer.

\subsection{Dependence of the magnon frequencies on the applied field}

\begin{figure}[h!]
 \centerline{$
 \begin{array}{c}
 \includegraphics[width=3.2in]{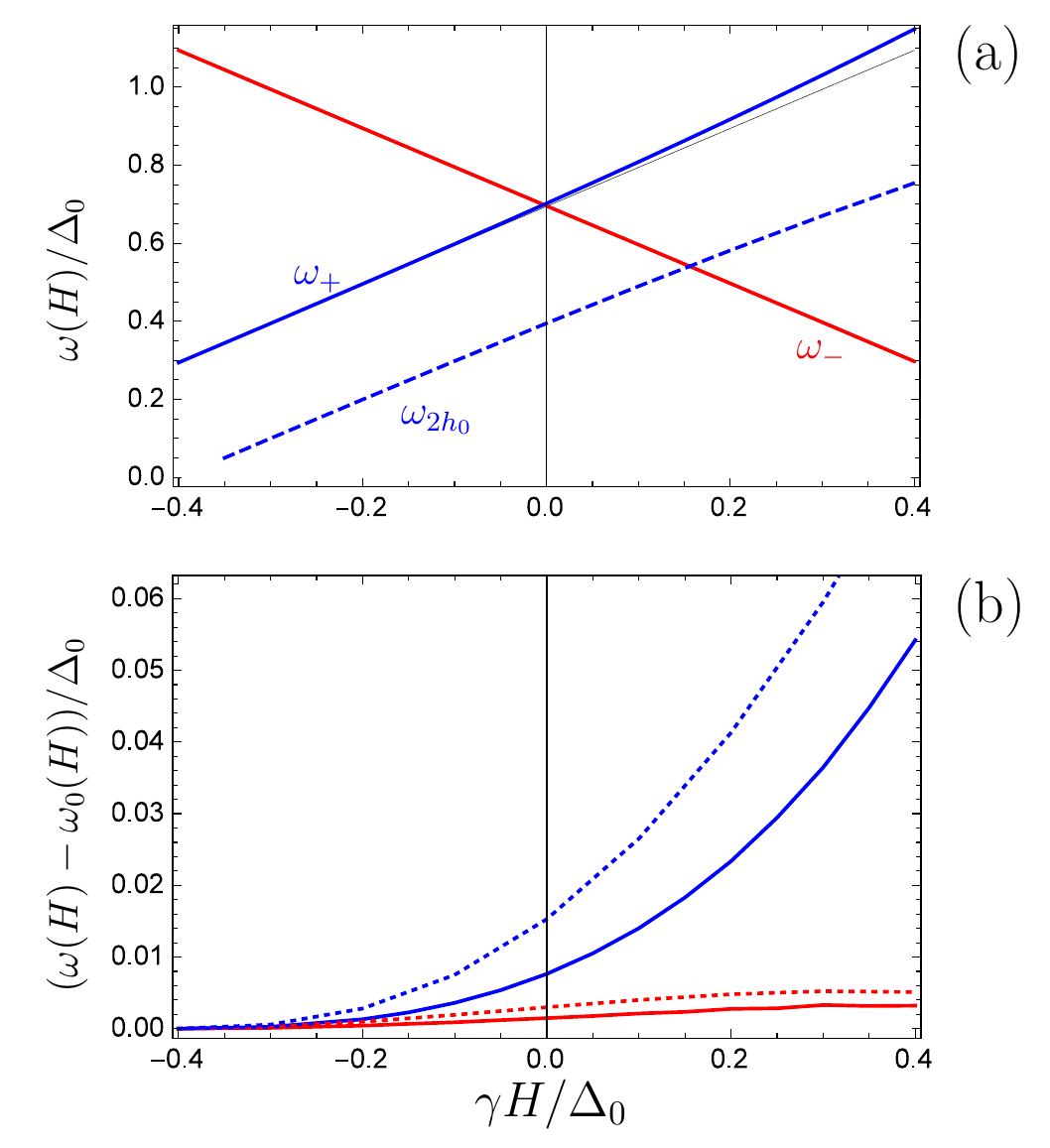} 
 \end{array}$}
 \caption{(a) Dependence of the modes $\omega_\pm(k=0)$ and $\omega_{2h_0}$ on the external magnetic field, applied along the easy axis. Grey lines correspond to $\omega_{0,\pm}(k=0)$. While $\omega_{0,+}$ is seen in panel (a), $\omega_{0,-}$ cannot be distinguished from the renormalized red line $\omega_-$. (b) Difference between the renormalized magnon frequencies of the AF/S(N) bilayer and $\omega_0$ as a function of the applied field. $k=0$. Solid lines represent the difference $\omega_+(H)-\omega_{0,+}(H)$ (blue) and $\omega_-(H)-\omega_{0,-}(H)$ (red) at $T=0.4\Delta_0$, and the dotted lines correspond to the same differences but plotted at $T=T_c$. $h_0=0.2\Delta_0$ for  both panels.}
\label{fig:magnetic} 
\end{figure}

Further, in Fig.~\ref{fig:magnetic} we present the dependence of the split magnon frequencies at $k=0$ on the external field $H$, applied along the easy-axis of the antiferromagnet. The typical linear field-induced splitting of both modes is shown in panel (a). As  already discussed above, the renormalization of the AF frequencies by the adjacent conductor is not clearly seen on the scale $\Delta_0$. For this reason in panel (b) we plot the difference between the magnon frequencies in the AF/S(N) bilayer and the corresponding magnon frequency of the bare antiferromagnet $\omega_\pm(k=0)-\omega_{0,\pm}(k=0)$ as a function of $H$. Here $\omega_{0,\pm}$ means the corresponding mode without the proximity to the metal. The solid lines represent low-temperature results calculated in the superconducting state at $T=0.4\Delta_0$, and the dotted lines are the high-temperature results at $T=T_c$, which actually correspond to the normal state. It is seen that the difference $\omega_\pm-\omega_{0,\pm}$ grows with temperature supporting the quasiparticle origin of the $k=0$ renormalization, discussed above. It is also important that $\omega_\pm-\omega_{0,\pm}$ is an asymmetric function of the applied field. This is obviously due to the presence of the effective exchange $h_0$ in the metal layer. In this case the applied field also contributes to the effective exchange as $h_0 \to h_0+\mu_B H$. The difference $\omega_\pm-\omega_{0,\pm}$ goes to zero at $\gamma H = -2h_0$. This provides another method of experimental measurement of $h_0$: the induced exchange is determined by the applied field when the difference between $\omega_+(k=0)$ in the superconducting and normal states of the metal layer disappears.

\section{Conclusions}
\label{conclusions}

In the present paper we have investigated the renormalization of the magnon modes in a thin-film antiferromagnetic insulator caused by the proximity to a thin metal layer, which can be normal or superconducting. The key point is the presence of an uncompensated magnetic moment at the AF/S interface, which induces an effective exchange field in the adjacent metal via the interface exchange interaction. The exchange field spin polarizes quasiparticles in the metal and induces spinful triplet Cooper pairs screening the magnon, which provide an additional dynamical contribution to the magnetic anisotropy and influence the magnon velocity. The renormalization results in the splitting of the AF magnon modes with no need to apply a magnetic field. The physical reason for this splitting is asymmetric coupling of two antiferromagnet sublattices to the metal. The proximity effect also leads to the appearance of additional modes in the magnon spectrum due to the hybridization between the magnons and electron paramagnetic resonance mode, which is caused by the exchange field induced in the metal by the antiferromagnet itself. 

The dependence of the renormalized magnon modes in the AF/S(N) bilayer on the applied magnetic field is also studied. It is proposed that measurements of the renormalized dispersion relations can provide direct information about the amplitude of the effective exchange field induced by the AF in the adjacent metal. Our study demonstrates the promise of synthetic thin-film hybrids for tuning and modifying the spin-wave dispersion in antiferromagnets by manipulating the temperature or thickness of the metallic layer. 

\begin{figure}[tbh!]
 \centerline{$
 \begin{array}{c}
 \includegraphics[width=3.2in]{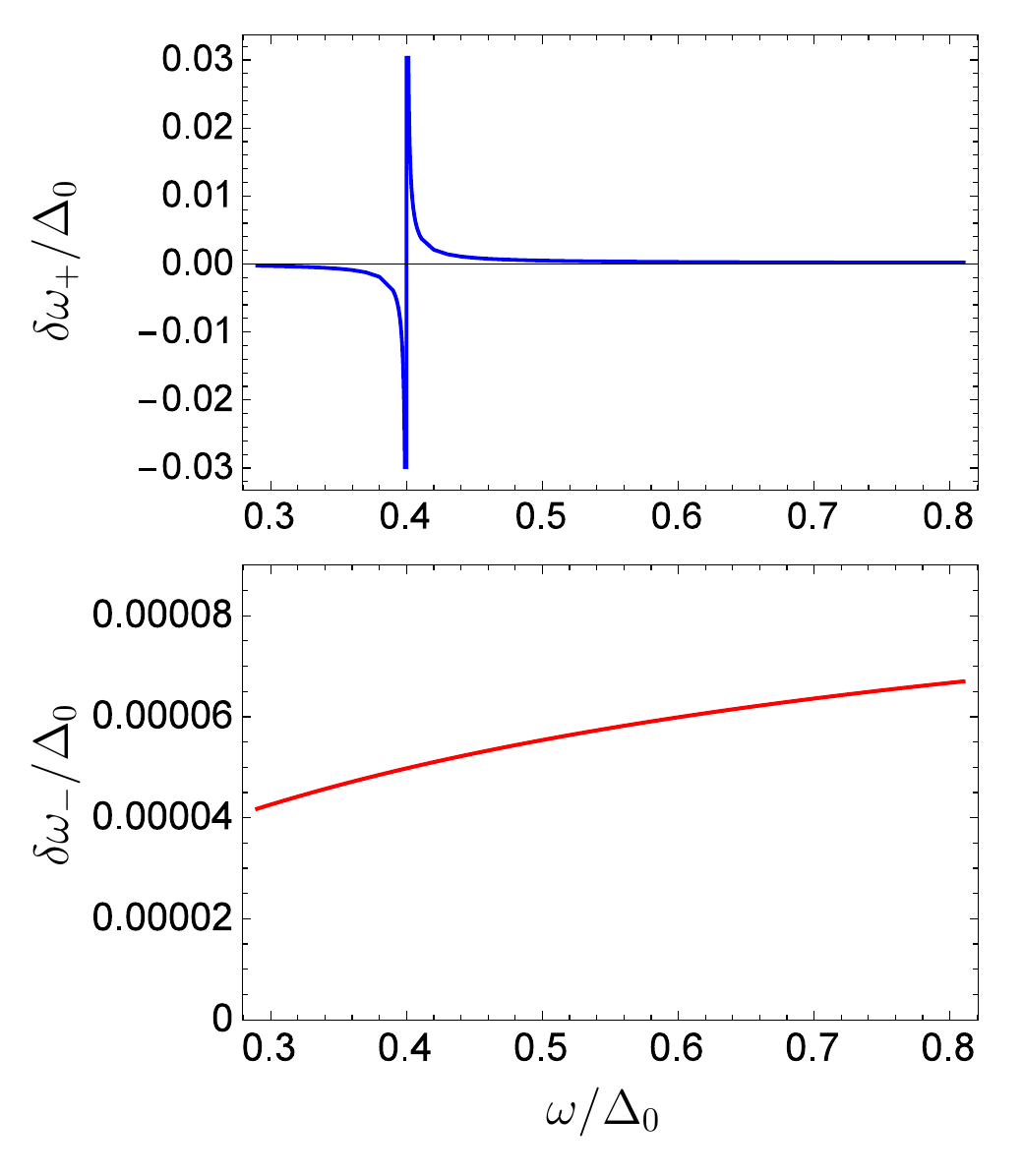} 
 \end{array}$}
 \caption{Top panel: $\delta \omega_+$ as a function of $\omega$. Bottom panel: $\delta \omega_-$  as a function of $\omega$. $h_0=0.2 \Delta_0$, $T=0.2 \Delta_0$, $kc/2 = 0.0003$.}
\label{fig:A1} 
\end{figure}

\section*{Acknowledgments}
We are grateful to Akashdeep Kamra for valuable discussions. The financial support from the Russian Science Foundation via the RSF project No.22-42-04408 is acknowledged.

\section*{Appendix}

Here we present the details of the numerical solution for the $\omega_\pm$ and $\omega_{2h_0}$ modes, presented in Figs.~\ref{fig:dispersions_1}, \ref{fig:h_cases}. The mode frequencies are obtained as solutions providing zero real part $Z$ of the determinant  of the matrix in Eq.~(\ref{eq:modes}), that is 
\begin{widetext}
\begin{eqnarray}
Z = {\rm Re}\left[det 
\left(
\begin{array}{cc}
-\omega - i \kappa \pm B \pm \delta \omega_\pm \pm i \omega \alpha - i \chi_\pm & \pm \gamma J_{\mathrm{AF}} \pm i \omega \alpha_c \\
\mp \gamma J_{\mathrm{AF}} \mp i \omega \alpha_c & -\omega - i \kappa \mp B \mp i \omega \alpha 
\end{array}
\right)\right] \approx 
\omega^2 - \omega_0^2 + \delta \omega_\pm (\mp \omega - B) = 0 .
\label{eq:Z}
\end{eqnarray}
\end{widetext}
Here we consider $H=0$. For this reason $B_+ = B_- = B$.
 
\begin{figure}[tbh!]
 \centerline{$
 \begin{array}{c}
 \includegraphics[width=3.2in]{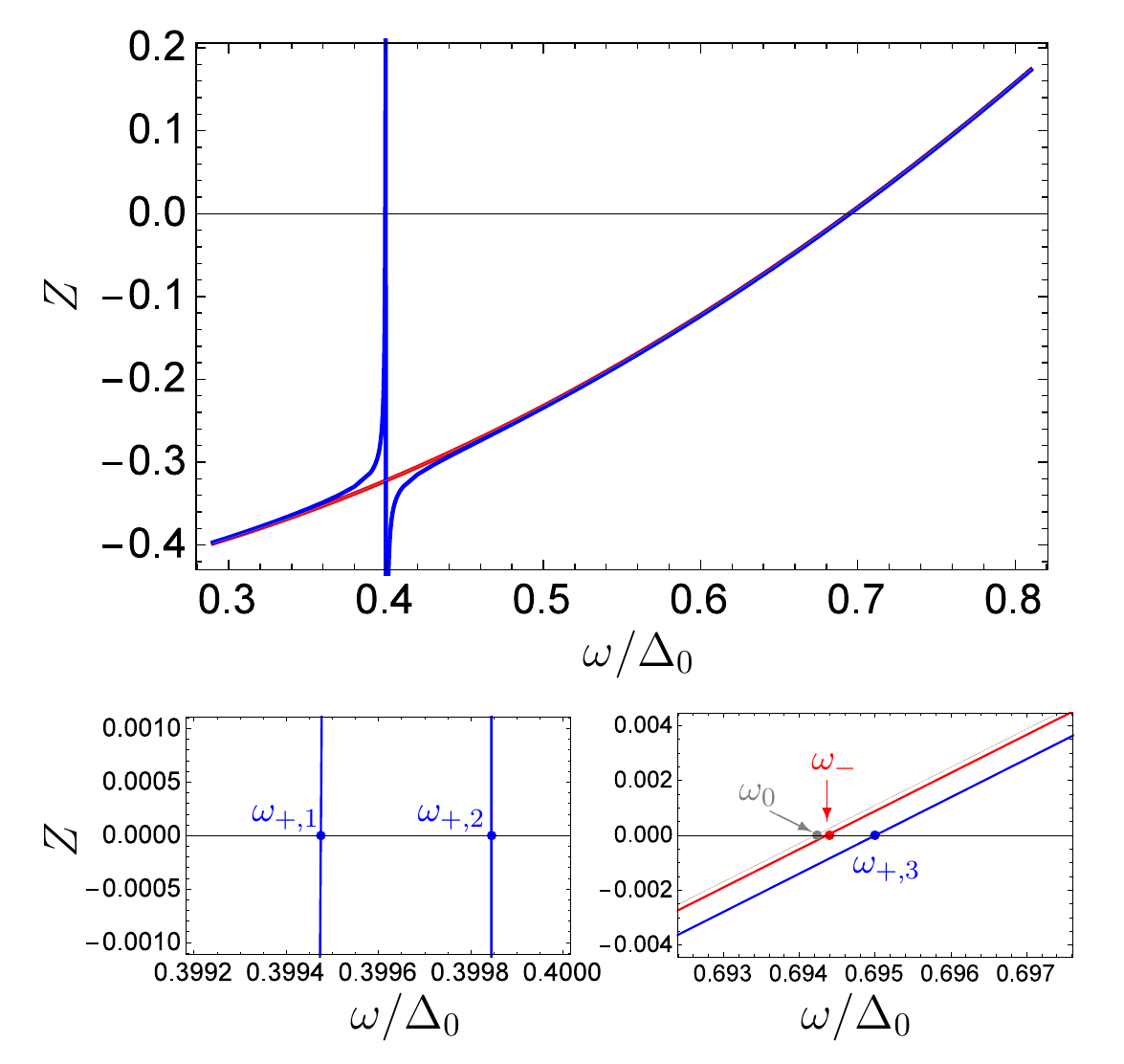} 
 \end{array}$}
 \caption{$Z(\delta \omega_+)$(blue) and $Z(\delta \omega_-)$(red) as functions of $\omega$. Bottom panels represent regions in the vicinity of the intersections of the curves with zero level on a larger scale. Three solutions of the equation $Z(\delta \omega_+)=0$ are marked by the blue points and the only solution of the equation $Z(\delta \omega_-)=0$ is marked by the red point. The mode frequency $\omega_0$ for the isolated antiferromagnet is marked by the grey point. The parameters correspond to Fig.~\ref{fig:A1}.}
\label{fig:A2} 
\end{figure}

\begin{figure}[tbh!]
 \centerline{$
 \begin{array}{c}
 \includegraphics[width=3.2in]{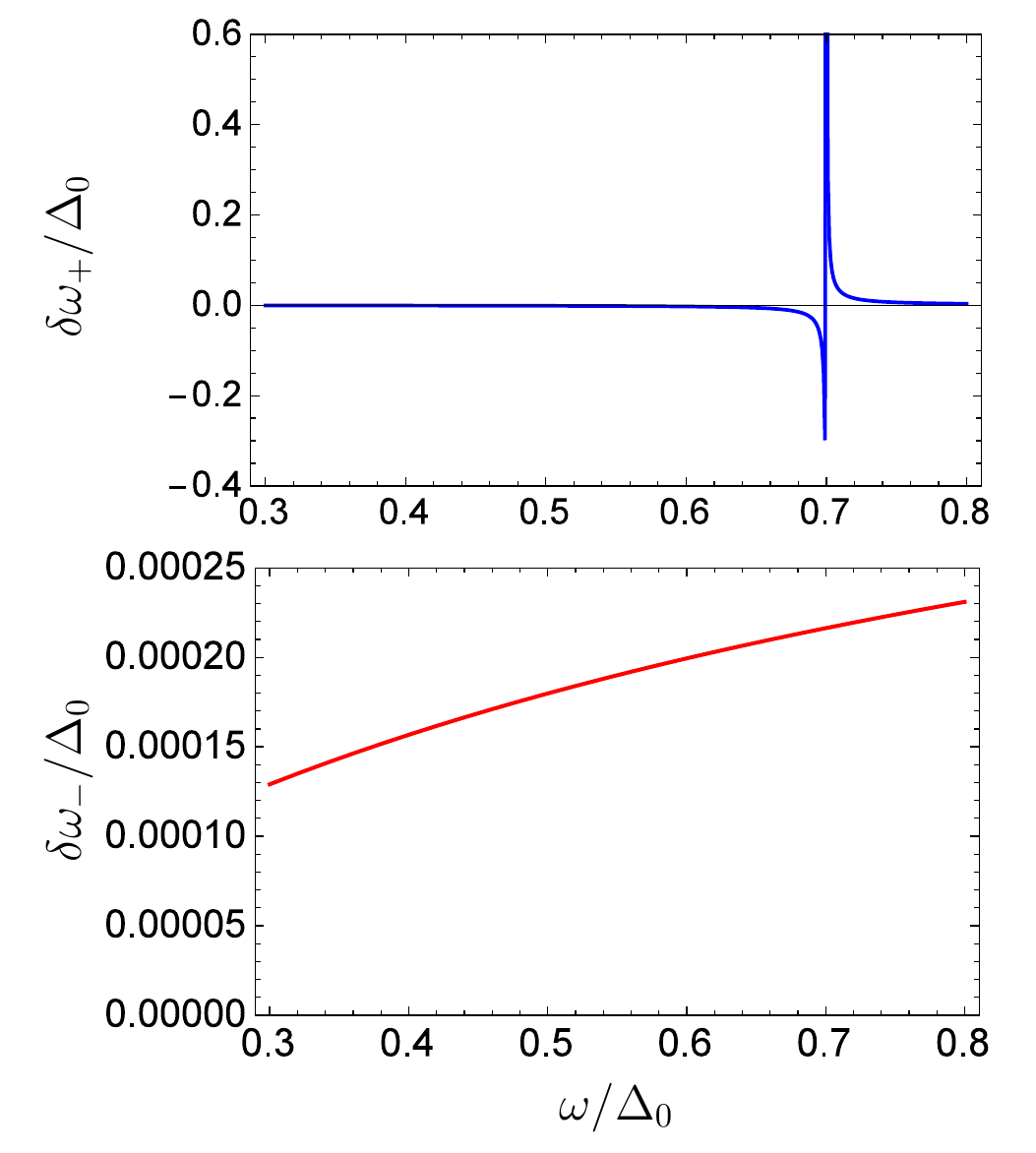} 
 \end{array}$}
 \caption{Top panel: $\delta \omega_+$  as a function of $\omega$. Bottom panel: $\delta \omega_-$  as a function of $\omega$. $h_0=0.35 \Delta_0$, other parameters are the same as in Fig.~\ref{fig:A1}.}
\label{fig:A3} 
\end{figure}

\begin{figure}[tbh!]
 \centerline{$
 \begin{array}{c}
 \includegraphics[width=3.2in]{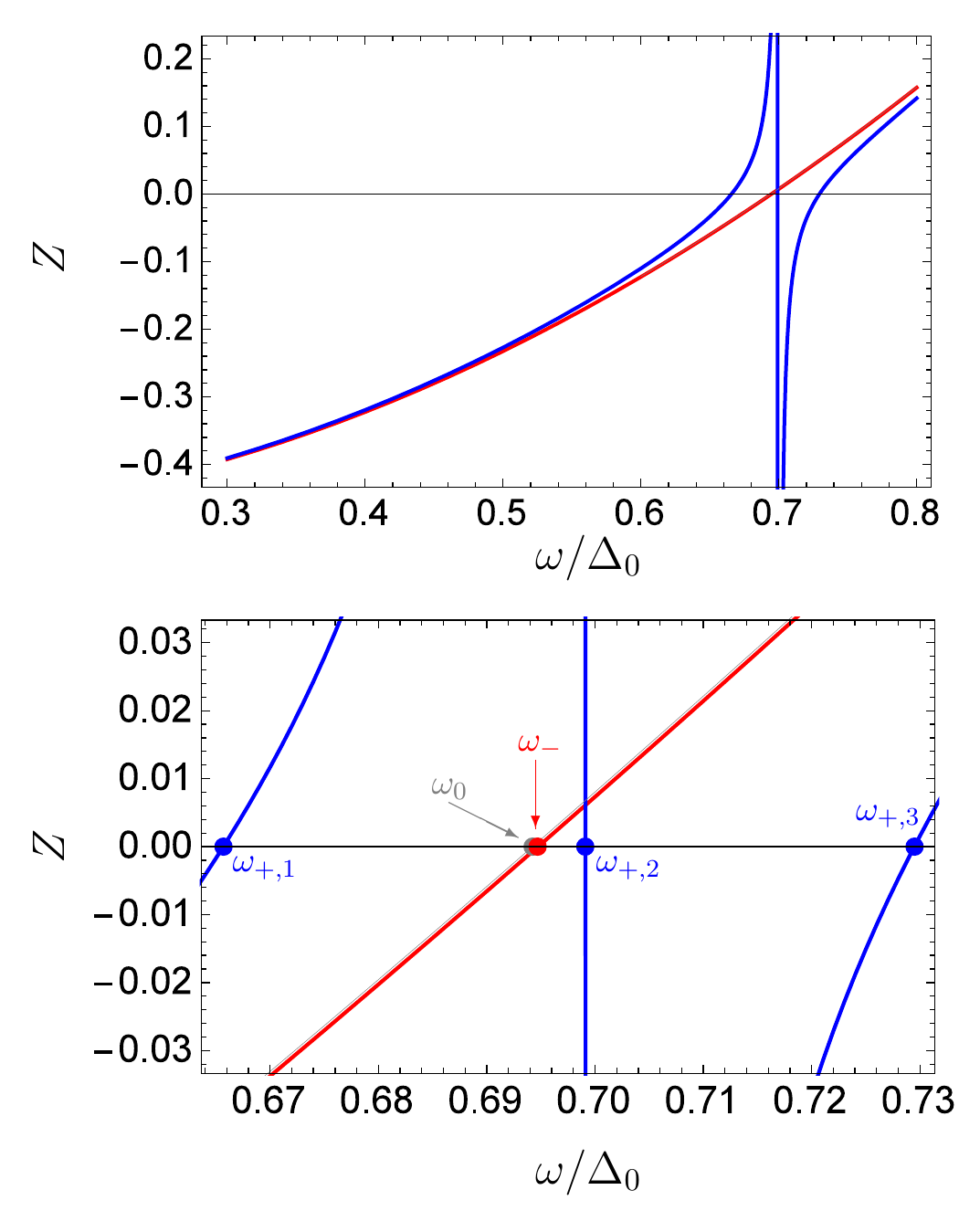} 
 \end{array}$}
 \caption{$Z(\delta \omega_+)$(blue) and $Z(\delta \omega_-)$(red) as functions of $\omega$. Bottom panel represents regions in the vicinity of the intersection of the curves with zero level on a larger scale. The parameters correspond to Fig.~\ref{fig:A3}. }
\label{fig:A4} 
\end{figure}

Eq.~(\ref{eq:Z}) is nonlinear due to the dependence $\delta \omega_\pm(\omega)$. The superconducting corrections $\delta \omega_\pm$ as functions of $\omega$ at $kc/2 \to 0$ are shown in Figs.~\ref{fig:A1}, \ref{fig:A3} for two different values of the effective exchange field $h_0=0.2\Delta_0$ and $h_0=0.35\Delta_0$, respectively. It is seen that $\omega_+$ manifests a resonant behavior at $\omega = 2h_0$, while $\omega_-$ has no resonance features. As  already discussed in the main text, this is because only the $\omega_+$-mode interacts with the electron paramagnetic resonance mode.

Figs.~\ref{fig:A2}, \ref{fig:A4} demonstrate the dependence $Z(\omega)$ for $h_0=0.2\Delta_0$ and $h_0=0.35\Delta_0$, respectively, taking into account the superconducting correction.  Blue curves correspond to $Z(\delta \omega_+)$, and red curves represent $Z(\delta \omega_-)$. Far from $\omega = 2h_0$ $Z(\delta \omega_+)$ almost coincides (on the scale of the figure) with $Z(\omega_0) = \omega^2 - \omega_0^2$, corresponding to the isolated antiferromagnet. $Z(\omega_0)$ is shown by grey lines in Figs.~\ref{fig:A2} and \ref{fig:A4}. $Z(\delta \omega_-)$ is visually indistinguishable from $Z (\omega_0)$ everywhere.

Intersections of the blue and red curves with the horizontal line $Z=0$ give the solutions $\omega_\pm$. The regions in the vicinity of the intersection points are shown on the bottom panels of Figs.~\ref{fig:A2}, \ref{fig:A4} on a larger scale. In Fig.~\ref{fig:A2} $\omega_{+,3}$ provides the renormalized value of the bare antiferromagnetic mode $\omega_0$ and is called by $\omega_+$ in the main text. $\omega_{+,1}$ is the additional mode caused by the interaction of the magnonic subsystem with the electron paramagnetic resonance mode and is called  $\omega_{2h_0}$ in the main text. $\omega_{+,2}$ is not discussed in the main text due to its very strong decay. Physically it means that we hardly can think about this solution as a real mode. We cut the lines if $\kappa/\Delta_0>0.3$.

The key difference between the considered cases $h_0=0.2\Delta_0$ and $h_0=0.35\Delta_0$ is that for the last case the resonance region $\omega \approx 2 h_0$ is very close to $\omega_0 \approx 0.694 \Delta_0$, which leads to much stronger renormalization of the bare antiferromagnetic mode and its overlapping with the additional modes. It is the reason for the very unusual behavior presented in Fig.~\ref{fig:h_cases}(a) of the main text. The solutions $\omega_{+,1}$ and $\omega_{+,3}$ in Fig.~\ref{fig:A4} correspond to two $\omega_+$ branches seen in Fig.~\ref{fig:h_cases}(a) at $k \to 0$. Analogously to the previous case $\omega_{+,2}$ is also strongly decaying.  For this reason the corresponding branch is not seen at small $k$ in Fig.~\ref{fig:h_cases}(a). 

Fig.~\ref{fig:A4} is plotted for the case $k \to 0$. However, the number of solutions of the equation $Z=0$ depends on $k$.  Upon increasing $k$ the decay rate of the solution $\omega_{+,2}$ decreases and this solution splits into three different solutions. As a result we have five branches of $\omega_+(k)$ in Fig.~\ref{fig:h_cases}(a). Upon further increase of $k$ solutions of the equation $Z=0$ continue to evolve, which leads to the fact that at first there are three branches and then only one.

For the curves, presented in Fig.~\ref{fig:h_cases}(b) and corresponding to $h=0.5 \Delta_0$ the intersection $\omega_0 \approx 2h_0$ occurs at large $k$, where $\delta \omega_+$ is strongly suppressed. For this reason in Fig.\ref{fig:h_cases}(b) we do not obtain any unusual features in the vicinity of this intersection.

\bibliography{AF-S}

\end{document}